\documentstyle[epsf]{cupconf}

\begin{document}

\title[Fragmentation  in Molecular Clouds]{Fragmentation in Molecular
Clouds: The Formation of a Stellar Cluster}
\author[R.~Klessen \& A.~Burkert]{R\ls A\ls L\ls F\ns K\ls L\ls E\ls S\ls S\ls E\ls N\ns \and\ns  A\ls N\ls D\ls R\ls E\ls A\ls S\ns  B\ls U\ls R\ls K\ls E\ls R\ls T }
\affiliation{$\!\!\!\!$Max-Planck-Institut f{\"u}r Astronomie, K{\"o}nigstuhl 17,
69117 Heidelberg, Germany}

\maketitle

\begin{abstract}
The isothermal gravitational collapse and fragmentation of a molecular
cloud region and the subsequent formation of a protostellar cluster is
investigated numerically. The clump mass spectrum which forms during
the fragmentation phase can be well approximated by a power law
distribution $dN/dM \propto M^{-1.5}$. In contrast, the mass spectrum
of protostellar cores that form in the centers of Jeans unstable
clumps and evolve through accretion and $N$-body interaction is best
described by a log-normal distribution. Assuming a star formation
efficiency of $\sim\!10\;\!$\%, it is in excellent agreement with the IMF of
multiple stellar systems.
\end{abstract}

\section{Introduction}
\label{sec:intro}

Understanding the processes leading to the formation of stars is one
of the fundamental challenges in astronomy and astrophysics.  However,
theoretical models considerably lag behind the recent observational
progress. The analytical description of the star formation process is
restricted to the collapse of isolated, idealized objects (Whitworth
\& Summers 1985). Much the same applies to numerical studies
(e.g.~Boss 1997, Burkert et al.~1997 and reference therein).  Previous
numerical models that treated cloud fragmentation on scales larger
than single, isolated clumps were strongly constrained by numerical
resolution.  Larson (1978), for example, used just 150 particles in an
SPH-like simulation.  Whitworth et al.~(1995) were the first who
addressed star formation in an entire cloud region using
high-resolution numerical models. However, they studied a different
problem: fragmentation and star formation in the shocked interface of
colliding molecular clumps.  While clump-clump interactions are
expected to be abundant in molecular clouds, the rapid formation of a
whole star cluster requires gravitational collapse on a size scale
which contains many clumps and dense filaments.

Here, we present a high-resolution numerical model describing the
dynamical evolution of an entire {\em region} embedded in the interior
of a molecular cloud. We follow the fragmentation into dense
protostellar cores which form a hierarchically structured cluster.

\section{Numerical Technique and Initial Condition}
\label{sec:num-technique-and-initial-cond}
To follow the time evolution of the system, we use smoothed particle
hydrodynamics (SPH: for a review see Monaghan 1992) which is
intrinsically Lagrangian and can resolve very high density
contrasts.  We adopt a standard description of artificial viscosity
(Monaghan \& Gingold 1983) with the parameters $\alpha_v = 1$ and
$\beta_v = 2$. The smoothing lengths are variable in space and time
such that the number of neighbors for each particle remains at
approximately fifty. The system is integrated in time using a second
order Runge-Kutta-Fehlberg scheme, allowing individual timesteps for
each particle. Once a highly condensed object has formed in the center
of a collapsing cloud fragment and has passed beyond a certain
density, we substitute it by a `sink' particle which then continues
to accrete material from its infalling gaseous envelope (Bate et al.~1995). By doing so we prevent the code time stepping from
becoming prohibitively small. This procedure implies that we cannot
describe the evolution of gas inside such a sink particle.  For a
detailed description of the physical processes inside a protostellar
core, i.e. its further collapse and fragmentation, a new simulation
just concentrating on this single object with the appropriate initial
conditions taken from the larger scale simulation would be necessary
(Burkert et al.~1998).
 
To achieve high computational speed, we have combined SPH with the
special purpose hardware device GRAPE (Ebisuzaki
et al. 1993), following the implementation described in detail by
Steinmetz (1996).  Since we wish to describe a region in the interior
of a molecular cloud, we have to prevent global collapse. Therefore, we
use periodic boundaries, applying the Ewald  method in an
PM-like scheme (Klessen 1997).

The structure of molecular clouds is very complex, consisting of a
hierarchy of clumps and filaments on all scales (e.g.~Blitz
1993). Many attempts have been made to identify the clump structure
and derive its properties (Stutzki \& G{\"u}sten 1990, Williams et
al.~1994).  We choose as starting conditions Gaussian random density
fluctuations with a power spectrum $P(k) \propto 1/k^N$ and $0 \le N
\le 3$. The fields are generated by applying the Zel'dovich (1970)
approximation to an originally homogeneous gas distribution: we
compute a hypothetical field of density fluctuations in Fourier space
and solve Poisson's equation to obtain the corresponding
self-consistent velocity field.  These velocities are then used to
advance the particles in one single big timestep $\delta t$. We
present simulations with $50\;000$ and $500\;000$ SPH particles,
respectively.

\begin{figure*}[htb]
\unitlength1.2cm
\begin{picture}(12,10)
\put( 0.0, 4.9){\epsfxsize=6.5cm \epsfbox{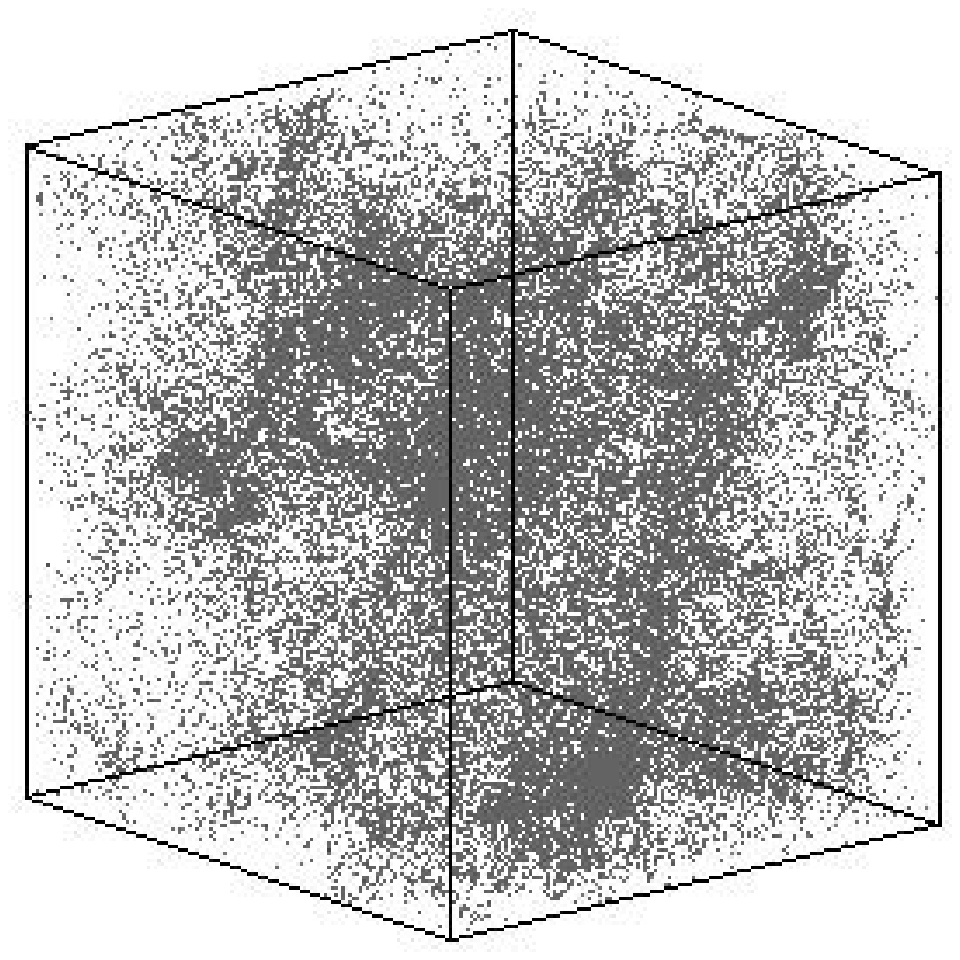}}
\put( 5.5, 4.9){\epsfxsize=6.5cm \epsfbox{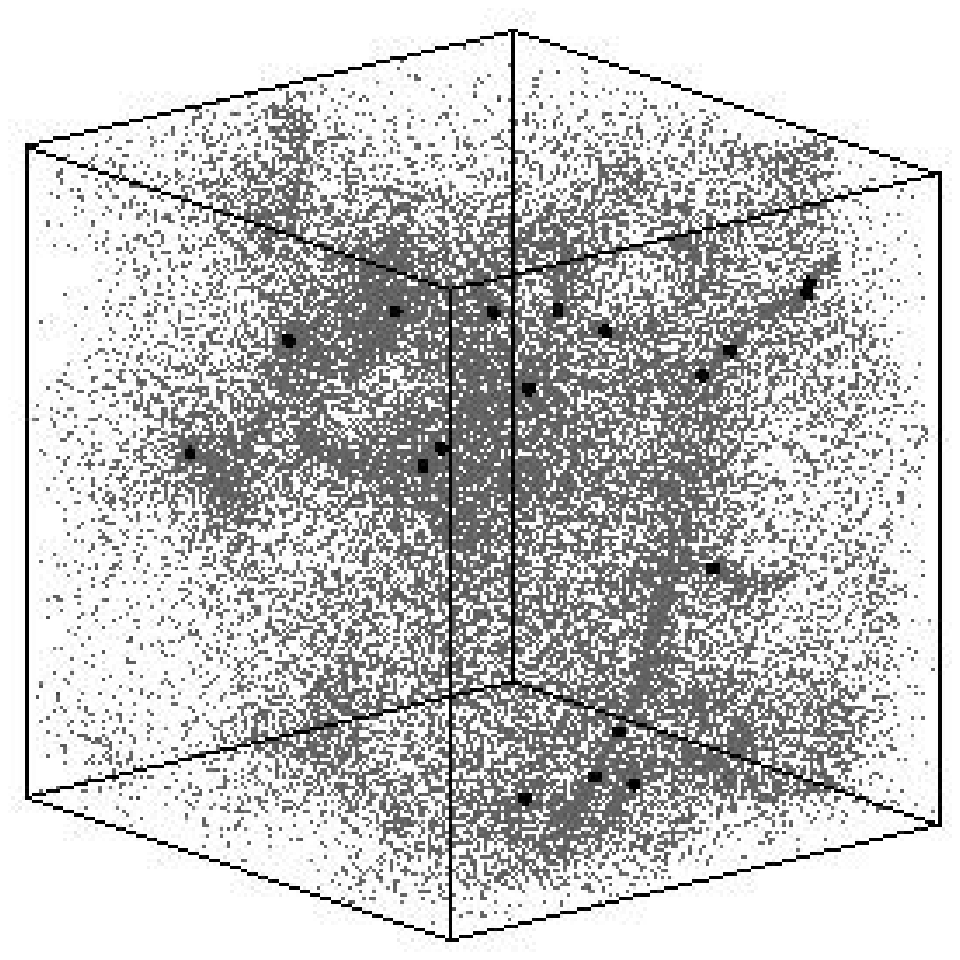}}
\put( 0.0,-0.0){\epsfxsize=6.5cm \epsfbox{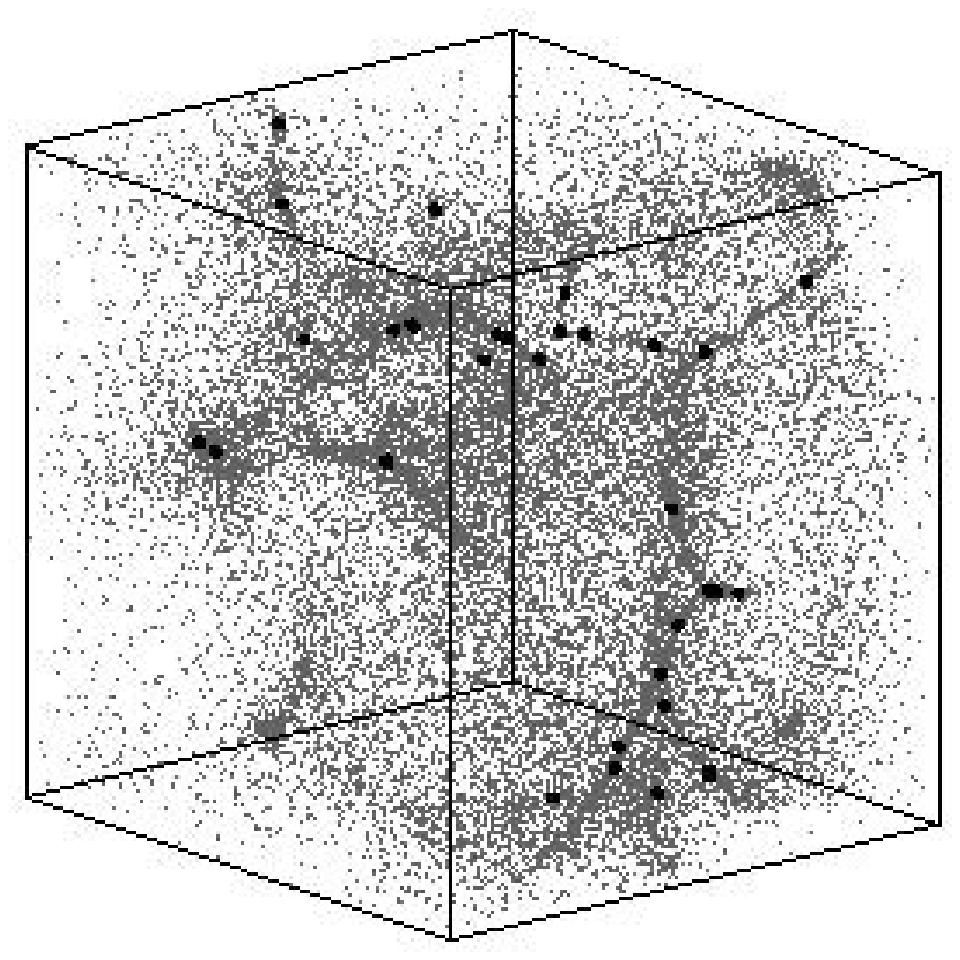}}
\put( 5.5,-0.0){\epsfxsize=6.5cm \epsfbox{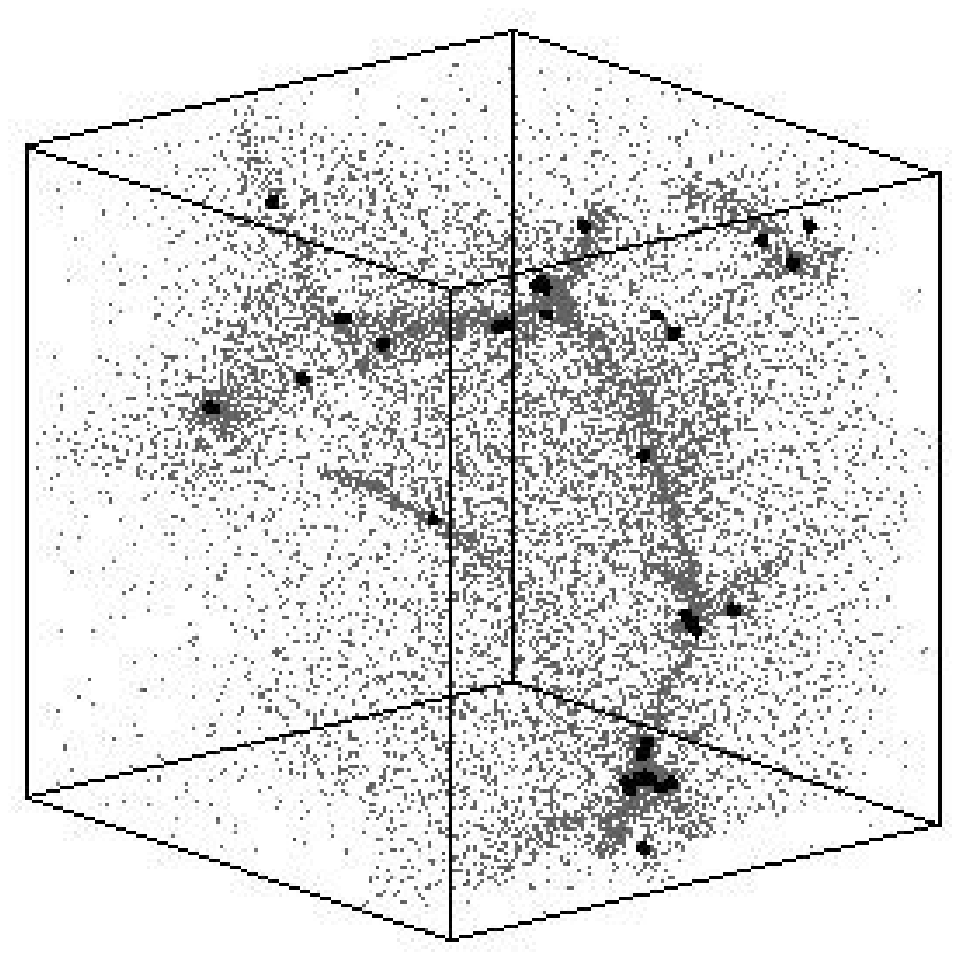}}
\put( 1.0, 5.2){\small $t=0.0$}
\put( 6.5, 5.2){\small $t=0.7$}
\put( 1.0, 0.3){\small $t=1.3$}
\put( 6.5, 0.3){\small $t=2.0$}
\end{picture}
\caption{\label{fig:3D-plots} Time evolution and fragmentation of a
region of 222 Jeans masses with initial Gaussian density fluctuations
with power law $P(k) \propto 1/k^2$.  Collapse sets in and soon forms
a cluster of highly-condensed cores, which continue to accrete from
the surrounding gas reservoir. At $t=0.7$ about 10\% of all the gas
mass is converted into "protostellar" cores (denoted by black
dots). At $t=1.3$ and $t=2.0$ these values are 30\% and 60\%,
respectively. Initially the cube contains $50\,000$ SPH particles.}
\end{figure*}
\section{A Case Study}
\label{sec:case-study}
As a case study, we present the time evolution of a region in the
interior of a molecular cloud with $P(k) \propto 1/k^2$ and containing
a total mass of 222 Jeans masses determined from the temperature and
mean density of the gas.  Figure~\ref{fig:3D-plots} depicts snapshots
of the system initially, and when 10, 30 and 60 per cent of the gas
mass has been accreted onto the protostellar cores.  Note that the
cube has to be seen periodically replicated in all directions.  At the
beginning, pressure smears out small scale features, whereas large
scale fluctuations start to collapse into filaments and knots. After
$t \!\approx\!  0.3$, the first highly-condensed cores form in the
centers of the most massive and densest Jeans unstable gas clumps and
are replaced by sink particles.  Soon clumps of lower mass and density
follow, altogether creating a hierarchically-structured cluster of
accreting protostellar cores. For a realistic timing extimate, the
Zel'dovich shift interval $\delta t = 2.0$ has to be taken into
account. In dimension-less time units, the free-fall time of the
isolated cube is $\tau_{\rm ff} = 1.4$.
 
\subsection{Scaling Properties}

The gas is isothermal. Hence, the calculations are scale free,
depending only on one parameter: the dimensionless temperature
$T\equiv E_{\rm int}/|E_{\rm pot}|$, which is defined as the ratio
between the internal and gravitational energy of the gas. The model
can thus be applied to star-forming regions with different physical
properties.  In the case of a dark cloud with mean density
$n(H_2)\simeq100\,$cm$^{-3}$ and a temperature $T\simeq10\,$K like
Taurus-Auriga, the computation corresponds to a cube of length
10$\,$pc and a total mass of $6\,300\,$M$_{\odot}$.  The dimensionless
time unit corresponds to $2.2 \times 10^6\,$yrs. For a high-mass
star-forming region like Orion with $n(H_2) \simeq 10^5\,$cm$^{-3}$
and $T\simeq 30\,$K these values scale to $0.5\,$pc and
$1\,000\,$M$_{\odot}$, respectively. The time scale is
$6.9\times10^4\,$yrs.

\subsection{The Importance of Dynamical Interaction and Competitive Accretion}

The location and the time at which protostellar cores form, is
determined by the dynamical evolution of their parental gas
clouds. Besides collapsing individually, clumps stream towards a
common center of attraction where they merge with each other or
undergo further fragmentation. The formation of dense cores in the
centers of clumps depends strongly on the relation between the
timescales for individual collapse, merging and sub-fragmentation.
Individual clumps may become Jeans unstable and start to collapse to
form a condensed core in their centers. When clumps merge, the larger new clump
continues to collapse, but contains now a {\em multiple} system of
cores in its center.  Now sharing a common environment, these cores
compete for the limited reservoir of gas in their surrounding (see
e.g.~Price \& Podsiadlowski 1995, Bonnell et al.~1997).  Furthermore,
the protostellar cores interact gravitationally with each other.
As in dense stellar clusters, close encounters lead to the formation
of unstable triple or higher order systems and alter the orbital
parameters of the cluster members. As a result, a considerable
fraction of ``protostellar'' cores get expelled from their parental
clump.  Suddenly bereft of the massive gas inflow from their
collapsing surrounding, they effectively stop accreting and their
final mass is determined. Ejected objects can travel quite far and
resemble the weak line T Tauri stars found via X-ray observation in
the vicinities of star-forming molecular clouds (e.g.~Neuh{\"a}user et
al.~1995).

\subsection{Mass Spectrum -- Implications for the IMF}

\begin{figure*}[t]
\unitlength0.85cm
\begin{picture}(13.6,5.8)
\put( 1.00, 2.70){\epsfxsize=2.7625cm \epsfbox{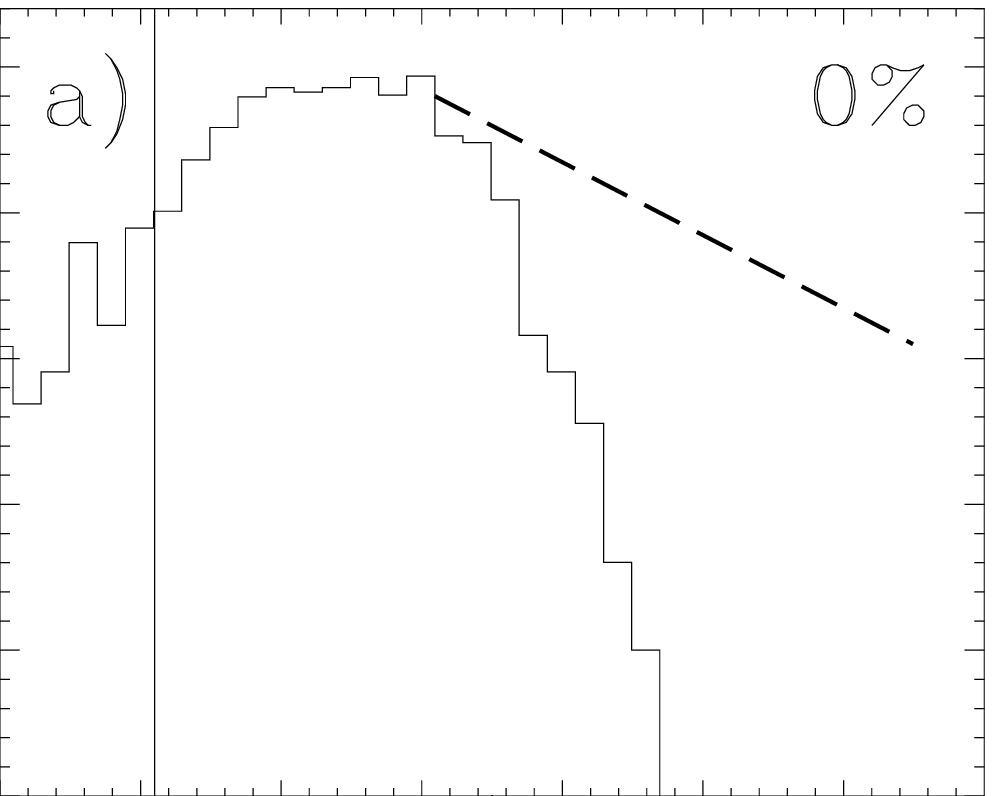}}
\put( 4.25, 2.70){\epsfxsize=2.7625cm \epsfbox{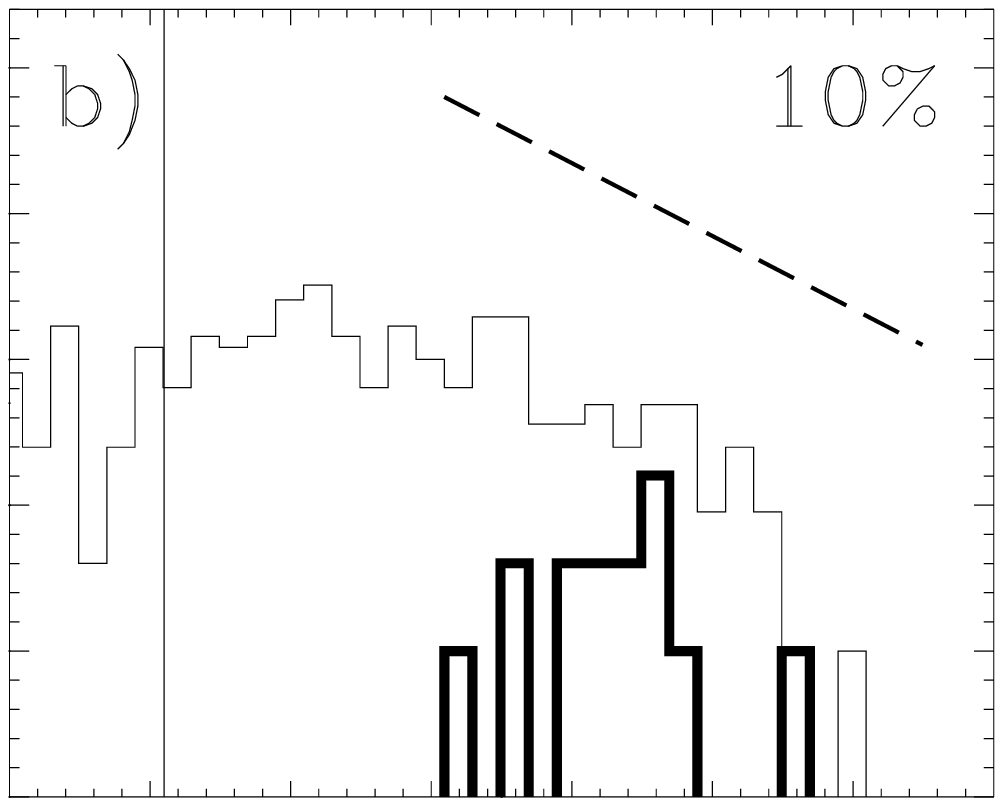}}
\put( 1.00, 0.10){\epsfxsize=2.7625cm \epsfbox{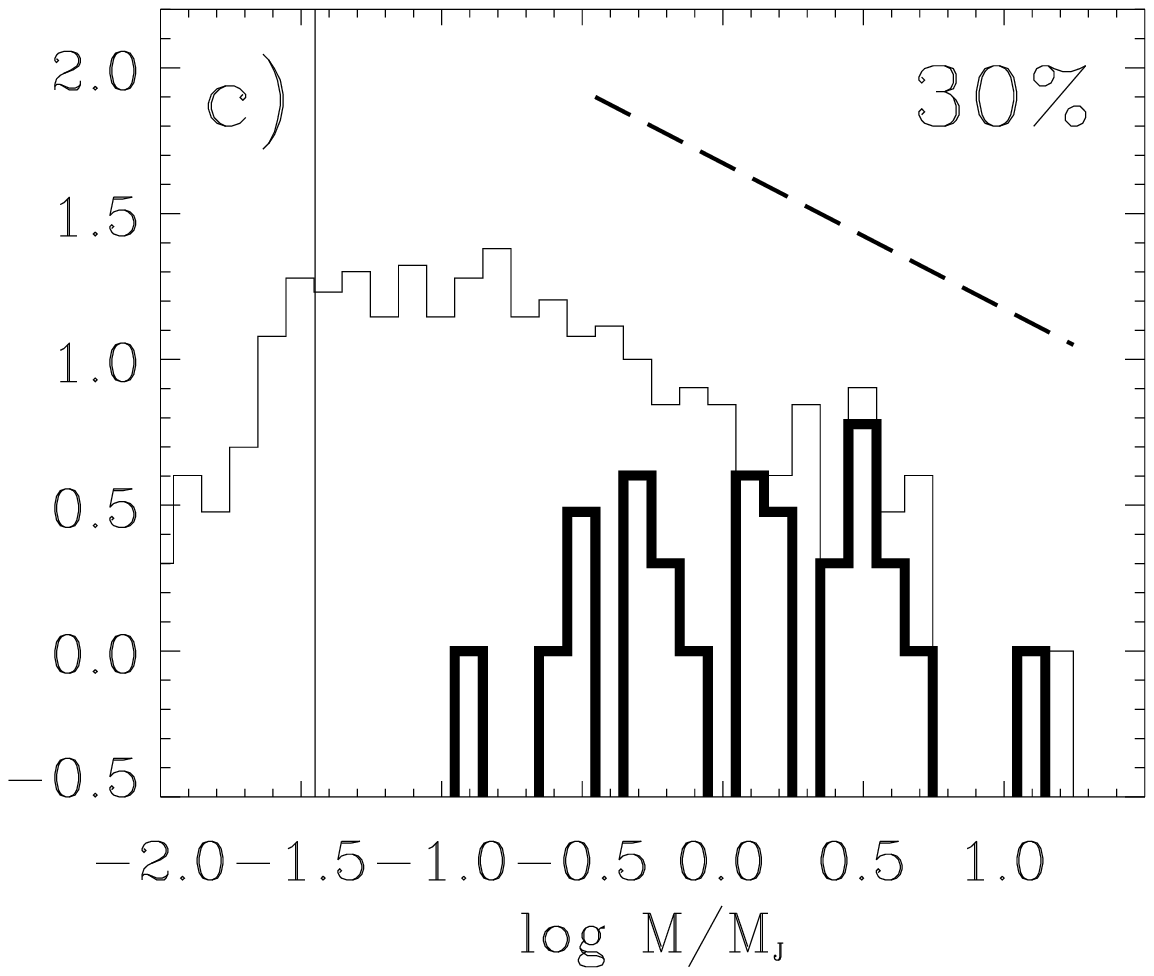}}
\put( 4.25, 0.10){\epsfxsize=2.7625cm \epsfbox{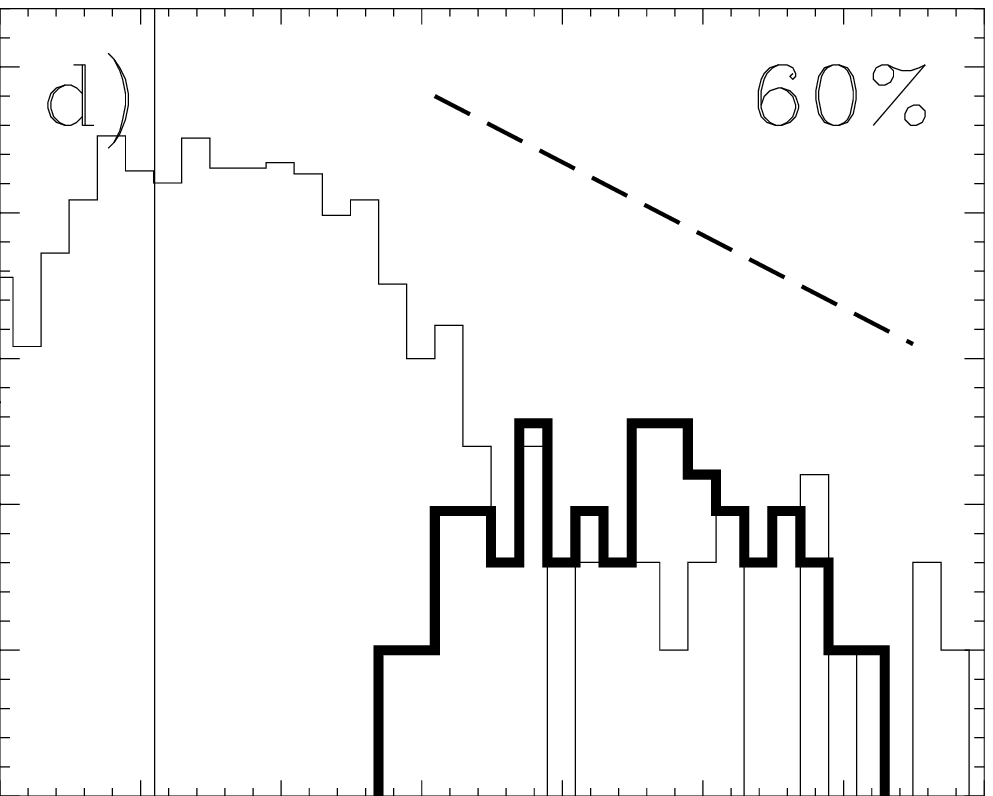}}
\put( 9.00, 0.10){\epsfxsize=5.525cm \epsfbox{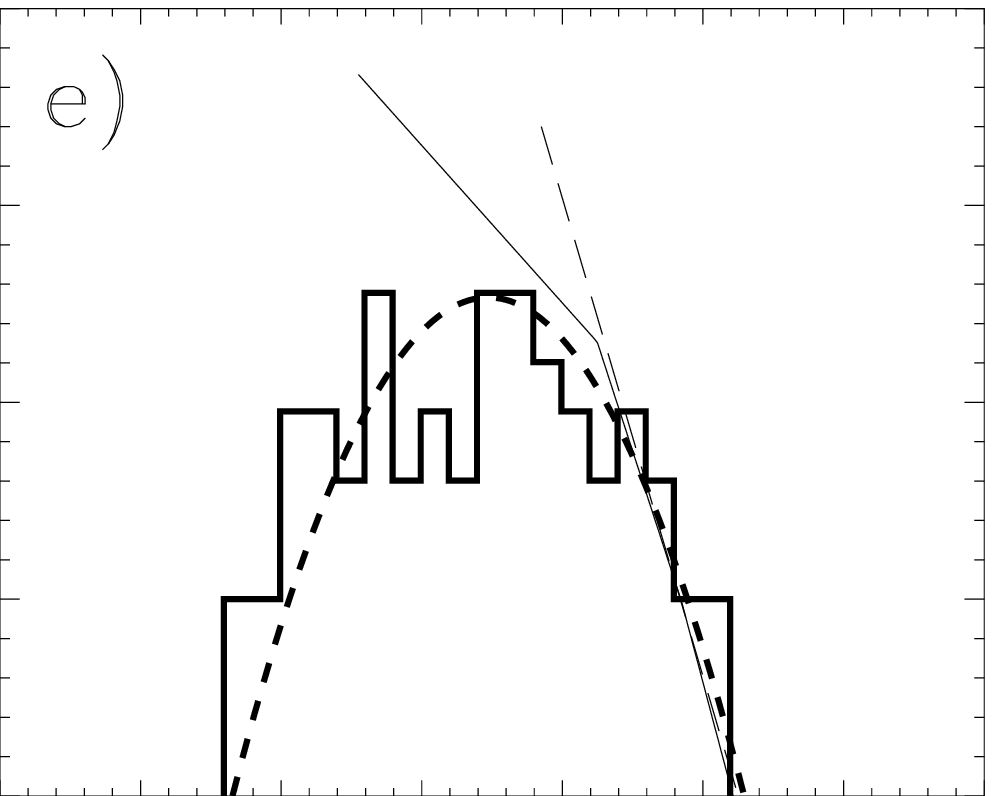}}
\end{picture}
\vspace*{0.4cm}
\caption{\label{fig:mass-distr} a) -- d) Mass distribution of gas
clumps (thin lines) and of protostellar cores (thick lines) at times
$t=0.0$, 0.7, 1.3 and 2.0 when 0\%, 10\%, 30\% and 60\% of the total
gas mass is condensed in cores, respectively. The vertical lines
indicate the resolution limit of the simulation with $500\,000$
particles (Bate \& Burkert 1997), and the dashed lines illustrate the observed clump mass
spectrum with $dN/dM \propto M^{-1.5}$ (Blitz 1993). e) Comparison of
the final core mass spectrum (thick line) with different
observationally based models for the IMF.  The thick dashed line
denotes the log-normal form for the IMF, uncorrected for binary stars
as proposed by Kroupa et al.~(1990). In order for the peaks of both
distributions to overlap, a core star formation efficiency of 10\% has
to be assumed.  The agreement in width is remarkable.  The multiple
power-law IMF, corrected for binary stars (Kroupa et al.~1993) is
shown by the thin solid line.  As reference, the thin dashed line
denotes the Salpeter (1955) IMF. Both are scaled to fit at the
high-mass end of the spectrum.  All masses are normalized to the overall
Jeans mass in the system.}
\end{figure*}

Figures~\ref{fig:mass-distr}a -- d describe the mass distribution of
identified gas clumps (thin lines) and of protostellar cores (thick
lines) that formed within unstable clumps in a simulation analogous to
Fig.~\ref{fig:3D-plots}, but with 10 times higher resolution.  To
identify individual clumps we have developed an algorithm similar to
the method described by Williams et al.~(1994), but based on the
framework of SPH.  As reference, we also plot the observed canonical
form for the clump mass spectrum, $dN/dM \propto M^{-1.5}$ (Blitz
1993), which has a slope of $-0.5$ when plotting $N$ versus $M$. Note
that our initial condition does not exhibit a clear power law clump
spectrum. The Zel'dovich approximation generates an overabundance of
small scale fluctuations. However, in the subsequent evolution, these
small clumps are immediately damped by pressure forces and non-linear
gravitational collapse begins to create a power-law like mass
spectrum.

A common feature in all our simulations is the broad mass spectrum of
protostellar cores which peaks slightly above the overall {\em Jeans
mass} of the system.  This is somewhat surprising, given the fact that
the evolution of individual cores is highly influenced by complex
dynamical processes.  In a statistical sense, the system
retains `knowledge' of its (initial) average properties. 

The present simulations cannot resolve subfragmentation in condensed
cores.  Since detailed simulations show that perturbed cores tend to
break up into multiple systems (e.g.~Burkert et al.~1997), we
can only determine the mass function of multiple systems.  Our
simulations predict an initial mass function with a {\em log-normal}
functional form. Figure~\ref{fig:mass-distr}e compares the results of
our calculations with the observed IMF for multiple systems from Kroupa et
al.~(1990).  Assuming a typical Jeans mass $M_{\rm J} \approx
1\,$M$_{\odot}$ and a star formation efficiency of individual cores of
~10$\:\!$\%, the agreement between the numerically-calculated mass
function and the observed IMF for multiple systems (thick dashed line;
from Kroupa et al.~1990) is excellent. For comparison, also the IMF
corrected for binary stars (Kroupa et a.~1993) is indicated as thin
solid line, together with the mass function from Salpeter (1955) as
thin dashed line.

\section{Discussion}
\label{sec:summary}

Large-scale collapse and fragmentation in molecular clouds leads to a
hierarchical cluster of condensed objects whose further dynamical
evolution is extremely complex. The agreement between the
numerically-calculated mass function and the observations strongly
suggests that gravitational fragmentation and accretion processes
dominate the origin of stellar masses.  The final mass distribution of
protostellar cores in isothermal models is a consequence of the
chaotic kinematical evolution during the accretion phase.  Our
simulations give evidence, that the star formation process can best be
understood in the frame work of a probabilistic theory. A sequence of
statistical events may naturally lead to a log-normal IMF (see
e.g.~Zinnecker 1984, Adams \& Fatuzzo 1996; also Price \&
Podsiadlowski 1995, Murray \& Lin 1996, Elmegreen 1997).

\end{document}